\documentclass[pra,twocolumn, superscriptaddress, aps, showpacs, amsmath, amssymb,  nofootinbib]{revtex4}
\usepackage{graphicx}
\usepackage{dcolumn}%align table columns on decimal point
\usepackage{bm}%bold math
\usepackage{bbm}%bbold math
\usepackage{amssymb}
\usepackage{amsfonts}
\usepackage{amsthm}
\usepackage{dsfont}

\begin{document}

\title{Random free fermions: An analytical example of eigenstate thermalization.}
\date{\today}

\author{Javier M. Mag\'an}\email{j.martinezmagan@uu.nl}
\affiliation{Institute for Theoretical Physics \emph{and} Center for Extreme Matter and Emergent Phenomena, \\
Utrecht University, 3508 TD Utrecht, The Netherlands \\}

%\maketitle

\begin{abstract}
Having analytical instances of the Eigenstate Thermalization Hypothesis (ETH) is of obvious interest, both for fundamental and applied reasons. This is generically a hard task, due to the belief that non-linear interactions are basic ingredients of the thermalization mechanism. In this article we proof that random gaussian free fermions satisfy ETH in the multiparticle sector, by analytically computing the correlations and entanglement entropies of the theory. With the explicit construction at hand, we finally comment on the differences between fully random Hamiltonians and random Gaussian systems, providing a physically motivated notion of randomness of the microscopic quantum state.
\end{abstract}

\maketitle

%\tableofcontents

%\newpage

\section{Introduction}

The problem of quantum thermalization can be stated as:
\begin{itemize}\label{question}
\item Given microscopic unitarity, how do Gibbs ensembles emerge?
\end{itemize}
If a many body quantum system is set in an initial pure state $\vert\psi (0)\rangle$, the evolved state $\vert\psi (t)\rangle=U(t)\vert\psi (0)\rangle$ is pure and time dependent, so it can never become a time independent mixed density matrix, such as Gibbs ensembles:
\begin{equation}\label{iloss}
U(t)\,\vert \psi (0)\rangle\langle\psi (0)\vert \,U^{\dagger}(t) \neq \rho^{\textrm{Gibbs}}\;,
\end{equation}
where $\rho^{\textrm{Gibbs}}$ is any Gibbs distribution. The dynamical emergence of Gibbs ensembles from unitary dynamics will be termed the problem of quantum thermalization. This problem is almost as old as quantum mechanics itself. For a selfcontained recent review, with an excellent account of references and historical rigor, see \cite{eisertreview}.

Although exact thermality cannot be attained within unitary evolution, we might still expect approximate thermality for the actual measurements done in experiments, the correlation functions of the theory. Given an observable $\mathcal{O}$ of the theory, it is written mathematically as:
\begin{equation}\label{thermalityt}
\langle\psi (t)\vert\,\mathcal{O}\,\vert\psi (t)\rangle = \textrm{Tr}(\rho^{\textrm{Gibbs}}\,\mathcal{O})\pm \textrm{error}\;.
\end{equation}
stating that the correct expectation value, as measured by the evolving quantum state is equal to a Gibbs ensemble average, up to some error. This error has to be negligible in the thermodynamic limit for the previous relation to be non trivial, and for the thermal expectation value to be a good approximation of the correct one.

To understand such behavior it was proposed in \cite{srednickisub} that the typical energy eigenstate $\vert E_{a}\rangle$ of the quantum system satisfy themselves the previous property:
\begin{equation}\label{thermalityE}
\langle E_{a}\vert\,\mathcal{O}\,\vert E_{a}\rangle = \textrm{Tr}(\rho^{\textrm{Gibbs}}\,\mathcal{O})\pm \textrm{error}\;,
\end{equation}
stating that the expectation value in the energy basis is well approximated by the thermal correlator. The previous phenomenom was coined Eigenstate Themalization in \cite{srednickisub}, but was earlier noticed in \cite{ETH}, in the context of quantum spin systems. At any rate, as argued in \cite{eisertreview}, both previous equations can be seen as more precise formulations of the original question~(\ref{question}), and as such they are hypothesis about the nature of quantum systems. Indeed, the second equation~(\ref{thermalityE}) is commonly known as the Eigenstate Thermalzation Hypothesis (ETH).

Relations~(\ref{thermalityt}) and~(\ref{thermalityE}) suggest various obvious questions:
\begin{itemize}
\item What type of systems display such behavior? Over which range of initial states and eigenstates do they display it?
\item What is $\rho^{\textrm{Gibbs}}$? Can we find the effective temperature $T=1/\beta$ from the pure state scenario?
\item What is the `error'?
\end{itemize}
Although we have a great deal of intuition about these questions through the connection between ETH and random matrices, to the author knowledge, see \cite{eisertreview}, no time independent Hamiltonian containing only few body interactions has been proven to satisfy relations~(\ref{thermalityt}) and~(\ref{thermalityE}). The difficulty to prove such behavior is related to the belief that non-linear interactions are fundamental to the quantum thermalization mechanism. 

The objective of this article is to study a family of quantum systems displaying ETH, satisfying relation~(\ref{thermalityE}). Contrary to common belief, these systems are `gaussian', containing only two-body interactions, see~(\ref{H}). We further compute the errors, both in correlations and entanglement entropies. These errors differ from the random matrix approximation, providing a physically motivated measure of `randomness' of the quantum state.

We want to remark that although these systems might seem unnatural from several points of view, they indeed serve as microscopic toy models of black physics. The underlying reason is that they are examples of systems with no locality structure whatsoever. The Hamiltonian connects every oscillator with every other democratically, a characteristic feature of black hole physics \cite{matrixpol,susskind,lashkari,sahakian2,sekino,subir,uswalks,usrandomfree}.

From a different perspective, since these systems display ETH, entanglement entropy is extensive across all bipartitions, even for `vacuum states' in the sectors with a large number of particles. This could be of interest for the type of questions raised in \cite{simone}. 

\section{Random free fermions, entanglement entropy and ETH}

The family of Hamiltonians we wish to study is the following:
\begin{equation}\label{H}
H=\alpha \sum\limits_{i=1}^{N}c_{i}^{\dagger}\,c_{i}+\eta\sum\limits_{i,j=1}^{N}c_{i}^{\dagger}\,V_{ij}\,c_{j}\;,
\end{equation}
where $\alpha$ and $\eta$ are parameters with energy dimensions, $c_{i}^{\dagger}$ and $c_{i}$ are creation and annihilation operators of spinless free fermions, with usual anticommutator relations, and the couplings $V_{ij}$ are independent random gaussian real numbers with zero mean and unit variance. The matrix $(\eta V)_{ij}\equiv\eta V_{ij}$ is therefore a random matrix taken from the GOE ensemble with deviation $\sigma_{V}=\eta$ (see \cite{tao} for a beautiful and modern treatment of random matrices).

The `free' nature of the model allows an exact solution via diagonalization of the matrix $V$. If $\psi^{a}$, for $a=1,\cdots , N$, are the eigenvectors of $V$ with eigenvalues $\epsilon_{a}$:
\begin{equation}
\sum\limits_{j=1}^{N}V_{ij}\,\psi^{a}_{j} = \epsilon_{a}\,\psi^{a}_{i}\;,
\end{equation}
then the Hamiltonian can be written as:
\begin{equation}
H = \sum\limits_{a=1}^{N}(\alpha +\epsilon_{a})\, d_{a}^{\dagger}\,d_{a}= \sum\limits_{a=1}^{N}E_{a}\, d_{a}^{\dagger}\,d_{a}\;,
\end{equation}
where $d_{a}^{\dagger}$ and $d_{a}$ are new creation and annihilation operators defined by:
\begin{equation}
d_{a}=\sum\limits_{i=1}^{N}\psi^{a}_{i}\,c_{i}\;.
\end{equation}
All eigenstates are constructed by choosing a set $\mathcal{A}$ of particles, and associate to it the following eigenstate:
\begin{equation}\label{eigenstate}
|\Psi^{N_{p}}\rangle =\prod\limits_{a\in \mathcal{A}}d_{a}^{\dagger}\,|0\rangle\;,
\end{equation}
where $|0\rangle$ is the state annihilated by all $d_{a}$, and $N_{p}$ is the number of particles in the state, a number which will play a key role below. Notice that there are $\binom{N}{N_{p}}$ independent states for a given $N_{p}$.

The objective of this article is to study the structure of these eigenstates, associated with~(\ref{H}). We will focus on the number operator, the two point correlation functions and entanglement entropies. Since the correlations and entanglement structure of the eigenstates are symmetrical under $N_{p}\rightarrow N-N_{p}$, a manifestation of particle-hole symmetry in this model, we will only focus on $N_{p}\leq N/2$.

To compute the correlations and entanglement entropies, we will make use of the theory of random matrices, which deals with the statistical properties of eigenvalues and eigenvectors of matrices such as $\eta V$ (see \cite{taovectors},\cite{tao} and \cite{haake} for an extense treatment of random matrices). In relation to the eigenvectors, the main assertion is that the orthogonal matrix of eigenvectors $(\psi^{1},\cdots ,\psi^{N} )$ is distributed according to the Haar measure on the orthogonal group $O(N)$. For our purposes, this means that the eigenvectors have independent and random  gaussian entries, up to normalization. Matemathically:
\begin{equation}\label{statvec}
[\psi^{a}_{i}]=0 \,\,\,\,\,\,\,\, [\psi^{a}_{i}\,\psi^{b}_{j}]=\frac{1}{N}\,\delta_{ab}\,\delta_{ij}\;,
\end{equation}
where $[p]$ denotes the average of the random variable $p$ over the matrix ensemble.

For the eigenvalues we will only need Wigner's semicircle law, accounting for the probability of having an eigenvalue equal to $\lambda$:
\begin{equation}\label{wigner}
P(\lambda)=\frac{2}{\pi^{2}R^{2}}\,\sqrt{R^{2}-\lambda^{2}}\;,
\end{equation}
where $R^{2}\equiv 4N\eta^{2}$, and where we remind that this law concerns the eigenvalues of $\eta\, V$, a matrix with deviation equal to $\eta$, see the Hamiltonian~(\ref{H}). To ensure a zero energy vacuum eigenstate of $H$ we assume $R=2\sqrt{N}\eta<\alpha$. Given~(\ref{wigner}), the first two moments for the eigenenergies $E_{a}$ are:
\begin{equation}\label{E}
[\sum\limits_{a\in \mathcal{A}}E_{a}]=N_{p}\,\alpha \,\,\,\,\,\,\,\, [(\sum\limits_{a\in \mathcal{A}}E_{a})^{2}]-[\sum\limits_{a\in \mathcal{A}}E_{a}]^{2}=N_{p}\,N\,\eta^{2}\;.
\end{equation}
With the previous statistical information about eigenvalues and eigenvectors we can now compute the correlation functions for the state~(\ref{eigenstate}) with $N_{p}$ particles:
\begin{eqnarray}
C_{ij}^{\Psi}&=&\langle\Psi^{N_{p}}|c_{i}^{\dagger}c_{j}|\Psi^{N_{p}}\rangle =\sum\limits_{a,b\,=\,1}^{N}\psi^{a}_{i}\psi^{b}_{j}\,\langle\Psi^{N_{p}}|d_{a}^{\dagger}d_{b}|\Psi^{N_{p}}\rangle= \nonumber \\ &=&\sum\limits_{a\in \mathcal{A}}\psi^{a}_{i}\psi^{a}_{j}\;,
\end{eqnarray} 
where the last sum just runs over the subset $\mathcal{A}$ of particles chosen. The previous correlator is itself a random variable, a functional of the random eigenvectors $\psi^{a}$. By using~(\ref{statvec}) we obtain:
\begin{equation}\label{purec}
[C_{ij}^{\Psi}]=\frac{N_{p}}{N}\,\delta_{ij} \,\,\,\,\,\,\,\, [C_{ij}^{\Psi}C_{kl}^{\Psi}]-[C_{ij}^{\Psi}]\,[C_{kl}^{\Psi}]=\frac{N_{p}}{N^{2}}\,\delta_{ik}\,\delta_{jl}\;.
\end{equation}
We thus can see the correlation matrix as the sum of a `thermal' part and an error, the error being a random matrix taken from the GOE ensemble with deviation $\sigma_{C}^{2}=\frac{N_{p}}{N^{2}}$. Notice that the `thermal' part just depends on the `macroscopic' parameter $N_{p}$, and cannot distinguish between the $\binom{N}{N_{p}}$ independent states in the corresponding $N_{p}$ sector. Also notice that the `thermal' part is a good approximation in the window $N_{p}\gg 1$. As might have been expected, and anticipating results, $N_{p}=N/2$ will correspond to the high temperature sector.

To proof ETH for this system, relation~(\ref{thermalityE}) described in the introduction, we need to compute the corresponding correlation matrix in the thermal ensemble. This is given by:
\begin{eqnarray}
C_{ij}^{\beta}&=&\frac{1}{Z}\textrm{Tr}(e^{-\beta H}\,c_{i}^{\dagger}\,c_{j})=\frac{1}{Z}\sum\limits_{a,b\,=\,1}^{N}\psi^{a}_{i}\,\psi^{b}_{j}\,\textrm{Tr}(e^{-\beta H}\,d_{a}^{\dagger}\,d_{b})=\nonumber \\ &=&\sum\limits_{a=1}^{N}\psi^{a}_{i}\,\psi^{a}_{j}\, n^{E_{a}}_{\beta}\;,
\end{eqnarray}
where $n^{E_{a}}_{\beta}=1/(e^{\beta E_{a}}+1)$ is the average number operator for a fermionic oscillator at temperature $T=1/\beta$, and $Z=\textrm{Tr}(e^{-\beta H})$ is the usual partition function. The thermal correlation matrix is again a random variable, due to the randomness of $H$. Since there is no correlation between eigenstates and eigenvalues, the mean and variance are given by:
\begin{equation}\label{thermalc}
[C_{ij}^{\beta}]=[n_{\beta}]\,\delta_{ij} \,\,\,\,\,\,\,\, [C_{ij}^{\beta}C_{kl}^{\beta}]-[C_{ij}^{\beta}]\,[C_{kl}^{\beta}]=\frac{[(n_{\beta})^{2}]}{N}\,\delta_{ik}\,\delta_{jl}\;.
\end{equation}
where:
\begin{equation}
[n_{\beta}]=\int\limits_{-R}^{R}\,P(\lambda)\,n^{\lambda}_{\beta}\,d\lambda \,\,\,\,\,\,\,\,\,\, [(n_{\beta})^{2}]=\int\limits_{-R}^{R}\,P(\lambda)\,(n^{\lambda}_{\beta})^{2}\,d\lambda\;.
\end{equation}
We did not succeed in analytically computing the averages for all $\beta$. In the large temperature limit, $\beta\alpha\rightarrow 0$, they read:
\begin{equation}
[n_{\beta}]=\frac{1}{2}-\frac{\alpha\beta}{4} \,\,\,\,\,\,\,\, [(n_{\beta})^{2}]=\frac{1}{4}-\frac{\alpha\beta}{4}\;,
\end{equation}
whereas in the low temperature limit, $\beta R\rightarrow\infty$, we have slightly more complicated expressions:
\begin{equation}\label{n}
[n_{\beta}]=\frac{2e^{-\alpha\beta}\,I_{1}(\beta R)}{\beta R}\rightarrow \sqrt{\frac{2}{\pi\beta R}}\frac{e^{-\beta \,(\alpha-R)}}{\beta R}\;,
\end{equation}
and
\begin{equation}\label{nn}
[(n_{\beta})^{2}]=\frac{e^{-2\beta\alpha}\,I_{1}(2\beta R)}{\beta R}\rightarrow\frac{e^{-2\beta \,(\alpha-R)}}{\beta R\sqrt{4\pi\beta R}}\;,
\end{equation}
where $I_{1}(x)$ is the modified Bessel function of the first kind. Since to ensure a zero energy vacuum eigenstate of $H$ we assumed $R=2\sqrt{N}\eta<\alpha$, from~(\ref{thermalc}),~(\ref{n}) and~(\ref{nn}) we conclude that the leading diagonal approximation of the thermal correlation matrix is valid at all temperatures.

Now we are ready to compare both correlation matrices, the exact eigenstate correlations~(\ref{purec}) versus the thermal~(\ref{thermalc}) ones, allowing us to arrive to the following conclusions:
\begin{itemize}
\item For $N_{p}\gg 1$, a pure eigenstate with $N_{p}$ particles can be effectively approximated by a Gibbs distribution at temperature $\beta$ satisfying $\frac{N_{p}}{N}=[n_{\beta}]$, the difference being subleading in the thermodynamic limit. Random free fermions then constitute an explicit analytical example of ETH~(\ref{thermalityE}), an example in which we know the effective temperature and the error size in terms of the microscopic parameters of the theory.
\item For $N_{p}\sim \mathcal{O}(1)$, approximating pure states by thermal ensembles is not a valid approximation.
\end{itemize}
These features can be made clearer by studying entanglement entropy. For gaussian systems, as shown in \cite{peschel}, one can compute the entanglement entropy of a given subsystem $A$ directly from the correlation matrix. More concretely, given a subsystem $A$ with $m$ degrees of freedom, knowledge of $C_{ij}^{\Psi}=\langle\Psi|c_{i}^{\dagger}c_{j}|\Psi\rangle$, where $i,j\in A$, allows for the computation of the entanglement entropy. It is given by:
\begin{equation}\label{ent}
S_{A}=-\sum\limits_{i=1}^{m}(\lambda_{i}\log \lambda_{i}+(1-\lambda_{i})\log (1-\lambda_{i}))\;,
\end{equation}
where $\lambda_{i}$, with $i=1,\cdots ,m$ are the $m$ eigenvalues of the matrix $C^{\Psi}$ in the given subsystem $A$. The proof relies only on the fact that the correlation matrix and the reduced density matrix share the same set of eigenvectors.

Although formula~(\ref{ent}) is fairly simple, one still need to compute the eigenvalues of $C^{\Psi}$, and this is not always possible analytically, even for one-dimensional systems, see \cite{peschel}. Indeed we were not able to compute the entanglement entropy for all $N_{p}$. We will compute it in the two standard limiting cases: the thermal regime, specified by $N_{p}\gg 1$, and for $N_{p}=1$, corresponding to the non-thermal phase.

Let us begin with the simpler case of having just one particle. The wave function is then given by:
\begin{equation}\label{single}
|\Psi_{a}^{1}\rangle =d^{\dagger}_{a}\,|0\rangle =\sum\limits_{i=1}^{N}\psi^{a}_{i}\,c_{i}^{\dagger}\,|0\rangle =\sum\limits_{i=1}^{N}\psi^{a}_{i}\,|i\rangle\;,
\end{equation}
where $|i\rangle\equiv c_{i}^{\dagger}\,|0\rangle$. This state is considered in \cite{usloc}, in relation to the Many-Body-Localized phase transition. Because the state is fully supported in the single particle sector, the entanglement of any subsystem $A$ can be expressed as:
\begin{equation}\label{entsingle}
S_{A}=-p_{A}\log p_{A}-(1-p_{A})\log (1-p_{A})\;,
\end{equation}
where $p_{A}=\sum\limits_{i\in A}|\psi_{i}|^{2}$, see \cite{usloc} for an explicit derivation of the previous formula. The average and variance of $p_{A}$ are $[p_{A}]=m/N$ and $\sigma_{p_{A}}^{2}=m/N^{2}$. Because the variance of the probabilities $p_{A}$ is small in comparison to the mean value, we Taylor expand to compute the average entanglement entropy, following \cite{usrandom} for the case of random QFT states. For the case at hand, the entanglement entropy of a subsystem $A$ of $m\leq N/2$ degrees of freedom in the state~(\ref{single}) is given by:
\begin{equation}\label{ent1}
[S_{m}^{\Psi^{1}}]=-\frac{m}{N}\log\frac{m}{N}-(1-\frac{m}{N})\log(1-\frac{m}{N})-\frac{1}{2(N-m)}\;,
\end{equation}
where we remark that on average does not depend on the particle type chosen, labeled by $a$, and that the last term in the previous equation is alway subleading in the thermodynamic limit. The previous formula for entanglement entropy is a monotonic growing function for $1\leq m\leq N/2$, with $S_{1}^{\Psi^{1}}\simeq \frac{\log N}{N}$ and $S_{N/2}^{\Psi^{1}}\simeq \log 2$. This is a very small amount of entanglement, and certainly not extensive with the number of sites $m$. Therefore it cannot be faithfully represented as a thermal entropy, signalling that the small $N_{p}$ sector does not satisfy ETH.

For the high $N_{p}$ sector we follow a novel approximate method developed in \cite{usrandom}, Appendix A. The key aspect to observe is that for $N_{p}\gg 1$ the correlator matrix~(\ref{purec}) is a diagonal matrix plus a random matrix with parametrically smaller entries. Writing $C^{\Psi}=\bar{C}^{\Psi}+\delta C^{\Psi}$, to first order the eigenvalues  of $C^{\Psi}$ are given by $\lambda_{i}\simeq \bar{\lambda}_{i}+\delta\lambda_{i} $, where $\bar{\lambda}_{i}=N_{p}/N$ and $\delta\lambda_{i}$ are the eigenvalues of a random matrix of size $m$ with deviation $\sigma_{C^{\Psi}}^{2}=\frac{N_{p}}{N^{2}}\,$, see~(\ref{purec}). Using Wigner's semicircle law~(\ref{wigner}), for a matrix of size $m$ with such variance, these eigenvalues satisfy $[\delta\lambda_{i}]=0$ and $[(\delta\lambda_{i})^{2}]=m\frac{N_{p}}{N^{2}}\,$. Therefore, for $m\lesssim N_{p}$, plugging these eigenvalues into~(\ref{ent}), Taylor expanding in $\delta\lambda_{i}$ and finally taking the average, we obtain the following formula for the average entanglement entropy of a subsystem of size $m\lesssim N_{p}$ in the multiparticle state $|\Psi^{{N_{p}}}\rangle$:
\begin{equation}\label{entmany}
[S_{m}^{\Psi^{N_{p}}}]= m\, S^{N_{p}}_{1}-\frac{m^{2}}{2(N-N_{p})}\;,
\end{equation}
where we have defined:
\begin{equation}
S^{N_{p}}_{1}=-\frac{N_{p}}{N}\log \frac{N_{p}}{N}-(1-\frac{N_{p}}{N})\log (1-\frac{N_{p}}{N})\;,
\end{equation}
corresponding to the thermal entropy per degree of freedom for a state with $N_{p}\gg 1$ particles. Notice that to change from $N_{p}$ to the effective temperature $T=1/\beta$ we just need to use $\frac{N_{p}}{N}=[n_{\beta}]$.

Given the explicit expressions~(\ref{ent1}) and~(\ref{entmany}) we conclude:
\begin{itemize}
\item Entanglement entropy is well approximated by the thermal entropy for subsystems with sizes smaller than the number of particles in the given eigenstate. For bigger subsystems we cannot state anything with certainty, but it is tempting to speculate a slower growth of entanglement entropy for $m\gtrsim N_{p}$. For $N_{p}\simeq N/2$ the thermal approximation is valid for every subsystem.
\item The random nature of the state seems to increase with the number of particles.
\end{itemize}

\section{Error scalings as randomness measures}\label{secIII}

Until now we have focused in the structural properties of the eigenstates of the Hamiltonian~(\ref{H}), in particular on their thermal nature. It is now fruitful to check the differences between traditional approaches to quantum chaos and ETH based on fully random Hamiltonians, and the exact solution we found. This will provide a physically motivated quantification of the `randomness' of the quantum state. 

Although this section is self-contained, for extensive reviews about the traditional approach to ETH and quantum thermalization based on `typicality' arguments and random matrices in the contexts of quantum information theory, condensed matter and black hole physics see \cite{eisertreview,vijayreview,harlowreview}. To proceed we first write the Hamiltonian~(\ref{H}) as:
\begin{equation}\label{Hchaotic}
H=H_{\textrm{free}}+\bar{\eta} \,H_{\textrm{int}}\;.
\end{equation}
In most cases in which $H_{\textrm{int}}$ contains non-linear interactions we cannot diagonalize $H$ exactly. The traditional approximation assumes that the Hamiltonian can be taken from one of the random matrix ensembles, see \cite{haake}. The approximation implies the famous chaotic spectra, and more concretely Wigner's semicircle law~(\ref{wigner}). More interestingly for the concerns of this article, it also implies ETH as we show below, see \cite{jain1,jain2,eisertreview,vijayreview,harlowreview,usrandom}.

Let us begin with the spectrum. Since the dimension of the subspace with $N_{p}$ particles is $\binom{N}{N_{p}}$, and assuming the entries of $H_{\textrm{int}}$ to be random gaussian variables with zero mean and unit variance, the eigenenergies satisfy:
\begin{equation}\label{ranener}
E_{a}^{\textrm{random}}=N_{p}\,\alpha\pm 2\,\sqrt{\binom{N}{N_{p}}}\,\bar{\eta}\;,
\end{equation}
where $a=1,\cdots , \binom{N}{N_{p}}$. The first two moments of the random approximation can be computed from Wigner's semicircle law:
\begin{equation}\label{ranE}
[E_{a}^{\textrm{random}}]=N_{p}\,\alpha \,\,\,\,\,\,\,\, [(E_{a}^{\textrm{random}})^{2}]-[E_{a}^{\textrm{random}}]^{2}=\binom{N}{N_{p}} \,\bar{\eta}^{2}\;.
\end{equation}
On the eigenvectors side, the assumption of a random Hamiltonian implies that the eigenvectors are random vectors in the corresponding subspace:
\begin{equation}
\vert E_{a}^{\textrm{random}}\rangle=\sum\limits_{i=1}^{\binom{N}{N_{p}}}\psi^{a}_{i}\,\vert i\rangle\;,
\end{equation}
where $\vert i\rangle$ is a basis in the $N_{p}$ subspace and:
\begin{equation}\label{ranvec}
[\psi^{a}_{i}]=0 \,\,\,\,\,\,\,\, [\psi^{a}_{i}\,\psi^{b}_{j}]=\frac{1}{\binom{N}{N_{p}}}\,\delta_{ab}\,\delta_{ij}\;.
\end{equation}
See \cite{usrandom} for a recent detailed treatment of these type of vectors. Using~(\ref{ranvec}), the statistical properties of $C_{ij}^{\textrm{r}}= \langle E_{a}^{\textrm{random}}\vert \,c_{i}^{\dagger}c_{j}\,\vert E_{a}^{\textrm{random}}\rangle$ are:
\begin{equation}\label{ranC}
[C_{ij}^{\textrm{r}}]=\frac{N_{p}}{N}\delta_{ij} \,\,\,\,\,\,\,\, [(C_{ij}^{\textrm{r}})^{2}]-[C_{ij}^{\textrm{r}}]^{2}=\frac{\binom{N-2}{N_{p}-1}}{\binom{N}{N_{p}}^{2}}\sim \mathcal{O}(\frac{1}{\binom{N}{N_{p}}})\;.
\end{equation}
Comparing formulas~(\ref{ranE}) and~(\ref{ranC}), corresponding to the random Hamiltonian approximation, with the exact solutions~(\ref{E}) and~(\ref{purec}), and with the thermal result~(\ref{thermalc}), we conclude that:
\begin{itemize}
\item Appropriately fixing $\bar{\eta}$ in terms of $\eta$, $N_{p}$ and $N$, the mean and variance of the eigenvalues is the same for the exact Hamiltonian~(\ref{H}) than for its random approximation.
\item The average correlation matrix $[C_{ij}]$ coincides for all cases, the differences lying on variances. These variances furnish good quantifiers of randomness in the quantum state. They subtly distinguish between macroscopically equal phases, such as eigenstates of fully random hamiltonians and eigenstates of random free fermions. The scaling properties of the errors in typical eigenstates seem a fruitful field to explore in the context of quantum thermalization.
\item The random Hamiltonian approximation seems valid for all subsystem sizes $m$ (notice that in the one particle sector is exact for all subsystems), while the Gibbs distribution seems to hold only for $m\lesssim N_{p}$.
\end{itemize}

\section{Conclusions}
In this article we studied aspects of the Hamiltonian~(\ref{H}), such as the spectral properties~(\ref{E}), correlation matrix~(\ref{purec}) and entanglement entropies,~(\ref{ent}) and~(\ref{entmany}). They can be expanded in $1/N$, where $N$ is the number of spinless fermions of the model. The leading term in this expansion is always the thermal result, given by~(\ref{thermalc}), the effective temperature being found in terms of the macroscopic parameter $N_{p}$ characterizing the sector. The family of Hamiltonians~(\ref{H}) thus satisfy~(\ref{thermalityE}), furnishing explicit examples of the Eigenstate Thermalization Hypothesis (ETH). The conclusion is remarkable, since it implies that ETH is \emph{typical} within the space of gaussian Hamiltonians. Since it is also typical in the full space of Hamiltonians \cite{eisertreview}, as proved by relation~(\ref{ranC}), it is tempting to conclude that it is typical for Hamiltonians with random $2,3\cdots N$ body interactions, such as the model presented in \cite{subir}. 

We found that the entanglement properties of big subsystems in these eigenstates are different from the thermal result. For a sector with $N_{p}\gg 1$ particles, we were able to prove thermality of entanglement entropies until subsystems of size of $\mathcal{O}(N_{p})$. For bigger subsytems we speculated with a slower growth of entanglement entropy, but otherwise further study is needed to unravelled its nature. These results might have impact in black hole physics, in which the present picture is that given by Page in \cite{page}, a picture in which thermality holds for every subsystem, and in which deviations from thermality are assumed to be those given by the random Hamiltonian approximation.

The last section was devoted to study the differences between the approach to quantum chaos and ETH based on random matrices, see the reviews \cite{eisertreview,vijayreview,harlowreview}, and the exact solution of the Hamiltonian~(\ref{H}). These differences lie in the deviations from the thermal result. The `errors' in equations~(\ref{thermalityt}) and~(\ref{thermalityE}) are good quantifiers of the `randomness' of the quantum state. Their scaling properties seem an exciting new route to explore in the context of quantum thermalization.

\section*{Acknowledgements}

It is a pleasure to thank Jose Barb\' on, Marcos Crichigno, Simone Paganelli and Stefan Vandoren for interesting discussions in closely related subjects, Hrachya Babujian for useful comments on the manuscript and the constructive criticisms of an anonymous referee regarding section~(\ref{secIII}). The author also wishes to thank all the participants of the workshop `Strongly coupled field theories for condensed matter and quantum information', held in Natal, Brasil, in which this work was presented, and in special the hospitality of the International Institute of Physics, located in Natal, in which part of this work has been developed. This work was supported by the Delta-Institute for Theoretical Physics (D-ITP) that is funded by the Dutch Ministry of Education, Culture and Science (OCW).

%{toc}{chapter}
%\addcontentsline{toc}{section}{Bibliography}

%bibliography{bibmutual} % texto.bib es el fichero donde est� salvada la bibliograf�a.
%\bibliographystyle{unsrt} % estilo de la bibliograf�a.

%\addcontentsline{toc}{section}{Bibliography}

%\bibliography{bibuscodif} % texto.bib es el fichero donde está salvada la bibliografía.
%\bibliographystyle{unsrt} % estilo de la bibliografía.
\end{document}